# Scaling of Metal-Clad InP Nanodisk Lasers: Optical Performance and Thermal Effects

*Preksha Tiwari\*, Pengyan Wen, Daniele Caimi, Svenja Mauthe, Noelia Vico Triviño, Marilyne Sousa, and, Kirsten E. Moselund*

IBM Research – Europe, Säumerstr. 4, 8803 Rüschlikon, Switzerland
*Correspondence to: tiw@zurich.ibm.com

**Keywords**
**nanolasers, metal-clad lasers, plasmonics, thermal simulations**

**Abstract**

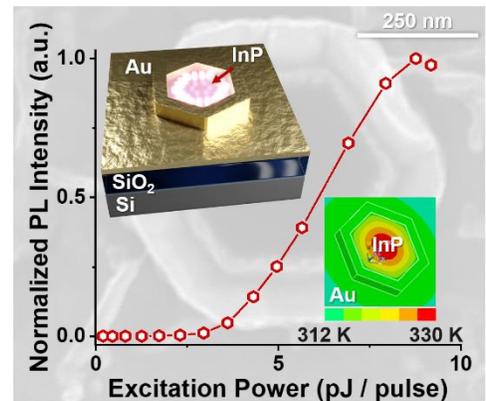

**A key component for optical on-chip communication is an efficient light source. However, to enable low energy per bit communication and local integration with Si CMOS, devices need to be further scaled down. In this work, we fabricate micro- and nanolasers of different shapes in InP by direct wafer bonding on Si. Metal-clad cavities have been proposed as means to scale dimensions beyond the diffraction limit of light by exploiting hybrid photonic-plasmonic modes. Here, we explore the size scalability of whispering-gallery mode light sources by cladding the sidewalls of the device with Au. The metal clad cavities demonstrate room temperature lasing upon optical excitation for Au-clad devices with InP diameters down to 300 nm, while the purely photonic counterparts show lasing only down to 500 nm. Numerical thermal simulations support the experimental findings and confirm an improved heat-sinking capability of the Au-clad devices, suggesting a reduction in device temperature of 473 K for the metal-clad InP nanodisk laser, compared to the one without Au. This would provide substantial performance benefits even in the absence of a hybrid photonic-plasmonic mode. These results give us insight into the benefits of metal-clad designs to downscale integrated lasers on Si.**

## Introduction

The continuous downscaling of electrical components has enabled a vast density of integrated circuits and their associated increase in performance and complexity. The last generations of transistors, however, have struggled to keep up with the device performance due to difficulties in scaling certain physical parameters like the gate dielectric beyond a given dimension [1]. Presently, limitations in chip performance are associated with resistive interconnects, which meet increasing difficulties in terms of energy consumption and cross talk [2, 3]. Therefore, there is interest in finding alternatives to electrical interconnects [4]. One method to not only achieve possible higher integration density but also provide increased bandwidth can be the use of optical interconnects [5], which are already widely employed for intra-chip communication. Silicon-on-insulator (SOI), provides an ideal platform for distributing optical signals on the chip with low loss and high density. Along with passive Si photonics, however, a full optical link requires the integration of detectors and light sources. For the distribution of a global signal like the clock, a single off-chip light source is in principle enough, and only the detectors need to be on the chip itself. To eventually achieve an integration comparable to that of electronics, however, scaled discrete emitters are required.

Efficient light sources in Si are challenging to realize because of their indirect bandgap [6 - 9]. Due to their tunable and direct bandgap and because a mature device processing technology is well established, the heterogeneous integration of III - V semiconductors on Si offers a promising path. For a competitive optical on-chip solution a low power consumption is crucial: It is suggested that a laser should have an energy below 10 fJ/bit [1, 10]. A small form-factor [11] is required to reduce the capacitance of devices as well as to enable dense integration. Conventional photonic nanolasers or integrated emitters may not be scaled beyond the diffraction limit of light which is given by $(\lambda_{em}/2n)^3$, where $\lambda_{em}$ is the emission wavelength and $n$ is the refractive index of the gain material. If the resonant cavity is scaled in multiple dimensions beyond this limit, light will no longer be confined. Photonic crystal (PhC) cavities have proven to be extremely efficient in the realization of low-power nanolasers [12 - 17]. While the central cavity might be very small, a substantial amount of area, however, is usually needed to create the two-dimensional PhC lattice.

The use of metal-clad cavities to scale the device dimension has gained interest throughout the past decade [18]. Surface plasmon polaritons (SPPs) can be described as a strong coupling between an electromagnetic wave and the collective electron oscillation of free electrons at a metal-dielectric interface. These interactions lead to a strong local confinement of the electromagnetic field with the drawback of increased absorption losses due to the presence of the metal. Recently, several concepts for the fabrication of such III-V based cavities have been presented based on nanowires [19 - 22], nanopan lasers [23], metallic coated disks [24], metallo-dielectric nanopatches [25], coaxial cavities [26], gap-plasmons [27, 28], waveguide-integrated plasmon nanocavities [29] or metal coated pillars [30 - 32] amongst others. In a device study [33] plasmonic semiconductor-on-metal lasers based on CdSe nanosheets on top of an Au / MgF$_2$ substrate were compared with purely photonic ones of same geometry, but without the metal. It was found that the plasmonic ones compare or even surpass the photonic devices in terms of threshold and cavity sizes in the limit of thin devices [33].

Whereas, all these approaches employ metal-assisted cavities to achieve promising results, in many cases the possible performance gain is not based solely on the presence of a plasmonic mode, but rather on other benefits derived from the presence of the metal. These may be manifold, spanning from increased reflectance, Purcell enhancement to acting as a local heat sink [29] for heat generated inside the gain medium. Often the photonic mode competes favourably with the plasmonic one down to very low dimensions [34] and it is generally difficult to assess the plasmonic nature of a mode. Selected studies investigated and compared the lasing behavior for devices with and without metal; like modulation bandwidth and energy efficiency calculations for III-V nanowires coated in metal on III-V substrates [11], experimental investigation of square cavities of optically active material placed upon a metal substrate [33], or design proposals for waveguide-integrated plasmon nanocavity sources with efficient emission transfer into waveguides, enhanced light-matter interactions, and thermal simulations on device cooling due to plasmonic metal pads on the active material [29]. While metallic substrates can provide excellent results in terms of fundamental studies because it is possible to obtain high-quality smooth metal surfaces, they are generally not suitable as basis for an integrated photonics platform.

In this work we explore the scalability of III-V nanodisk lasers on Si and the potential benefits of a metal cavity. We investigate different geometries (circular, square and hexagonal) of InP nanodisk resonators fabricated by a combination of direct wafer bonding on Si and etching. Micro- and nanodisk cavities generally support whispering gallery modes (WGM) which tend to concentrate the electro-magnetic field along the periphery of the cavity. We characterize the devices before and after coating the sidewalls with Au to explore the lasing behavior with and without a metal cavity. We deposit the metal with standard micro-fabrication techniques without specifically optimizing the process for plasmonics. We assess the fabricated devices in terms of downscaling limits, required threshold power for lasing and evaluate the lifetime of different cavities. Thermal simulations using ANSYS are applied to gain deeper insight on the impact of thermal effects in the devices. We observe strikingly different temperature profiles for metal-clad cavities as compared to those without the metal because we benefit from the heat sinking capability of the Au clad devices. This is the reason why the downscaled Au-clad cavities demonstrated here show superior performance than the cavities without the metal.

**Materials and Fabrication**

Direct wafer bonding is a mature technique for the integration of III-Vs on Si. It has been employed to fabricate III-V lasers on Si [35, 36], and in our group we are exploring both, wafer bonding as well as direct monolithic growth to bring III-Vs onto silicon [37 - 40]. While monolithic integration might be the preferred long-term option, here we work with wafer bonding because it more readily allows to explore different cavity shapes.

Fig. 1 schematically shows the different steps involved in fabrication. Fig. 2a4 – a8 show scanning electron microscope (SEM) images corresponding to steps 4 to 8 of the fabrication process, respectively. 300 nm thick InP is grown on a sacrificial wafer using a stop-layer, bonded on a silicon wafer, and subsequently the III-V substrate is removed, leaving a thin layer of III-V on insulator on silicon. Cavities of different shapes are defined by e-beam lithography using hydrogen silsesquioxane (HSQ) and are subsequently dry etched into the material. The HSQ, which turns into an oxide under e-beam exposure, is left on top to protect the InP in the subsequent steps, and also because this provides for a more symmetric dielectric environment for the optical modes with silicon dioxide ($SiO_2$) both above and below. Aluminium oxide ($Al_2O_3$) is deposited using atomic layer deposition. For the cavities without metal we use a 3 nm thick layer for surface passivation, whereas for the Au-clad devices we deposited 8 nm $Al_2O_3$ to reduce the optical losses associated with the metal. In a next step some of the devices are cladded with Au. The metal on top of the cavity is removed by Ar-ion milling since the devices will be illuminated from the top.

Fig. 2b shows a bright field scanning transmission electron (BF-STEM) cross section image of a device where one can observe the InP cavity with smooth top and bottom interfaces. The InP top facet is still covered by HSQ, and the sidewalls with the $Al_2O_3$ and Au. Fig. 2c is a zoomed BF-STEM image over the $InP/Al_2O_3/Au$ interface, where we observe some undulations from the individual etch cycles. We expect this sidewall roughness to be more important as we scale down the geometries, and even more so in the case of a hybrid photonic-plasmonic mode as this will confine the mode closer to the interface. In fact, a reduced surface roughness is one of the potential benefits of direct monolithic integration [37].

We choose the fabrication of squares, disks and hexagons in order to examine the impact of the device geometry on the lasing behavior. While disks and squares are favorable choices for nanodisks, a hexagonal shape is interesting in terms of future monolithic integration due to the intrinsic hexagonal crystal shape of monolithically grown InP on Si(111). The intrinsic hexagonal shape of monolithic InP has already been exploited for light emission in previous work from our group [37]. We investigate dimensions ranging in cross-section from 1 μm down to 150 nm. In the following, the cross-section width of the InP structure excluding the thickness of the metal cladding is used to compare the different devices.

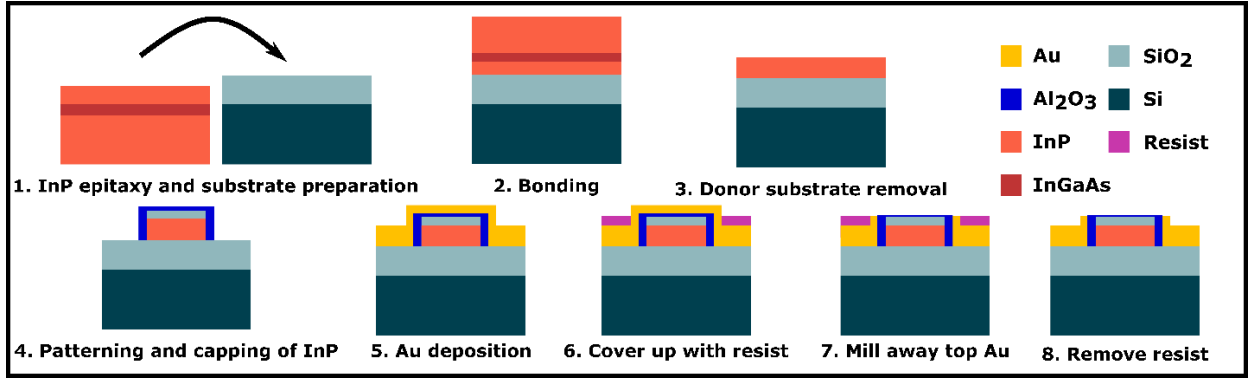

*Figure 1. Schematic overview of the device fabrication process: 1. InP is epitaxially grown on a donor wafer and bonded on the target wafer (2.). 3. The donor wafer is removed after bonding. 4. Cavities are dry etched into the InP using an HSQ hard mask and capped with $Al_2O_3$ after the patterning. 5. Au is deposited by a combination of angled evaporation and sputtering. 6. The HSQ is maintained at the top facet of the device as protection during the subsequent Ar-ion milling of the Au. (7.). 8. Excess resist is removed. The layers are not drawn to scale.*

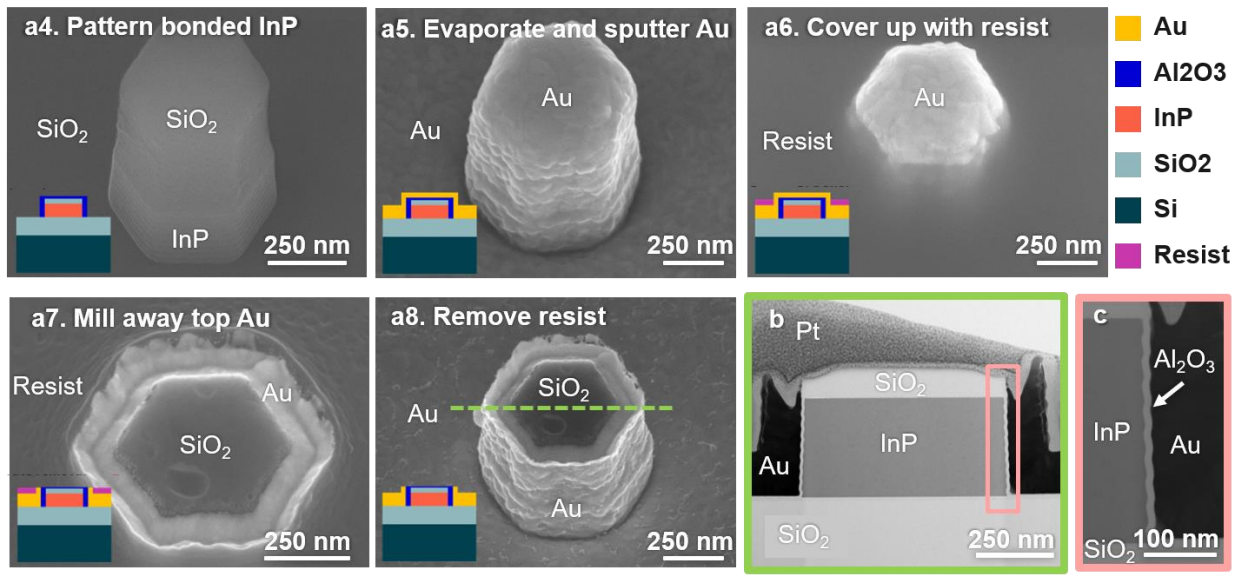

*Figure 2. (a) SEM images of the different fabrication steps starting from patterned InP devices and schematics of the layer stack. A cross section of the Au-clad device ((b), green frame) is taken at the green dashed line (see a8). A BF-STEM image ((c), rose frame) of the cross section shows the conformal Au coverage at the interface between InP, $Al_2O_3$ and Au. The etching cycles for the patterning of the InP lead to some sidewall roughness.*

**Thermal Simulations**

The metal cladding is known to introduce optical loss but also provides an efficient heat sink from the InP active region where the optical mode is concentrated. The goal of the thermal simulations is therefore to investigate the impact of the metal cladding on the temperature increase in the cavity for different geometries. An increase in temperature will lead to increased loss due to phonon-scattering as well as to a change of the refractive index. Thermal simulations are carried out using ANSYS 3D steady-state calculation. Fig. 3(a) and (b) shows the temperature distribution of different device structures with and without Au, respectively. From left to the right they show the 3D temperature distribution, cross-section temperature distribution, heat flux and the model. Fig. 3(c) shows the highest temperature increase as a function of the square diameter. We use a gaussian distributed heat generation load in the InP layer. This matches our pump-laser spot size of 1 μm in our experiments which will be described in the next section. As depicted in Fig. 3(c), the highest temperature increase of the InP layer reaches a maximum around a diameter of 800 nm for the square laser without Au, while it is 700 nm for a disk and a hexagon without Au. The variations between different geometries is due to the slightly varying volume for a fixed diameter.

We note here that the following temperature values should be taken as an upper boundary for several reasons: Firstly, for the thermal simulation we consider the average power, as opposed to shorter but higher amplitude pulsed pump peaks. Secondly, the power chosen corresponds to a rough average estimate of the threshold power. In the physical devices not all that power will be absorbed in the InP layer – a fraction will be reflected from various interfaces and a fraction will pass through and be absorbed in the silicon substrate. Nevertheless, the trend is the same, and in Fig. 3(c) there is a dramatic difference in the temperature increase in the two cases. The highest temperature extracted for the square laser without Au is 805 K, while it is only 325 K for the one with Au. In the structure without Au, the heat removal, mainly through the 2 μm thick $SiO_2$ box on the substrate, is quite inefficient. The amount of heat transferred to the substrate depends on the heat transfer cross-sectional area $A$, which is proportional to the diameter $D$ squared. Therefore, the gaussian distributed heat generation increases sharply with $D$, leading to a significant temperature increase in the InP layer as long as $D$ is small compared to the illuminated area. When $D$ approaches the pumping laser spot size, the heat generation is saturated, while $A$ keeps increasing, leading to a maximum highest temperature in the InP layer. A thinner bottom dielectric would naturally improve the heat removal in the non-metal-clad case, but a certain thickness of at least 500 nm or more is required for efficient optical isolation. This limits the practical tunability of this parameter. The laser with Au cladding on the other hand has an extremely efficient heat removal from the InP active region, since Au has a very high thermal conductivity and the presence of the thin $Al_2O_3$ is not an important barrier to thermal transport. The area of the Au is much larger compared to the InP square cross-section area in the thermal simulation model. In this structure, most of the generated heat tends to spread in-plane to the Au layer, which acts as a large heat sink, and then transfer downwards to the underlying box and substrate. When the InP square diameter is small, the heat transferred to the underlying layers is mainly dependent on the cross-section area of Au and the highest temperature of the InP layer only increases slowly with the increase of diameter. As the diameter increases further, an increasing fraction of heat will tend to directly transfer to the $SiO_2$ layer, especially in the center of the device, resulting in an increasing dependence on the diameter. Nevertheless, the temperature increase in the metal-clad cavities is substantially less than in the purely dielectric cavities, about 25 K versus 498 K for the largest square devices respectively.

In a practical integrated solution one would most likely only wish to have metal directly on the sidewalls, our simulations of this case (not shown here) show that this would as expected reduce the heat-sinking capability, but we would end up with a maximum temperature roughly half-way in between the two cases illustrated here.

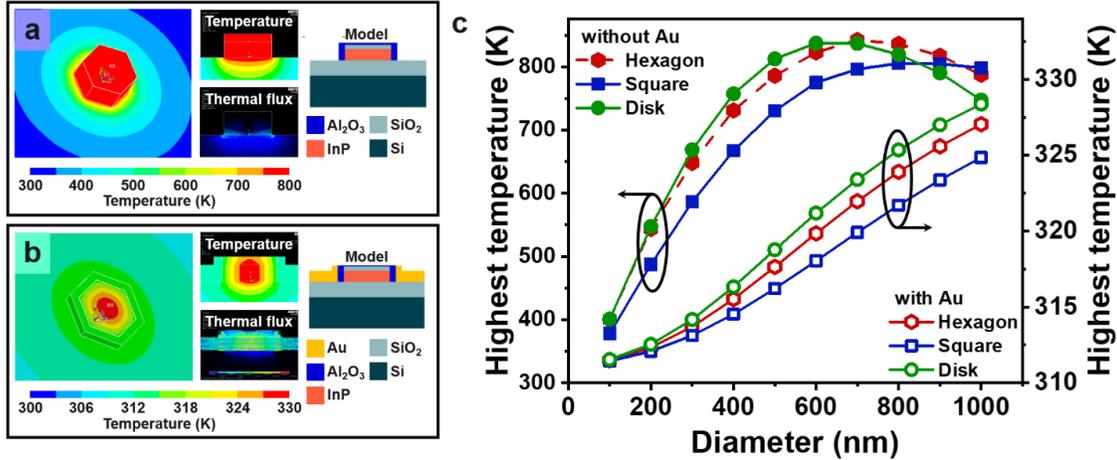

*Figure 3. Thermal simulations of the temperature distribution of InP lasers assuming constant pumping of the same average power as used in the experiment. (a) without Au and (b) with Au. (c) Highest temperature dependence on cavity diameter for squares (blue, filled), disks (green, filled) and hexagons (red, filled) without Au (left axis) and for squares (blue, empty), disks (green, empty) and hexagons (red, empty) with Au (right axis).*

**Optical micro-photoluminescence spectroscopy**

For the optical characterization we use a ps-pulsed supercontinuum laser with a repetition rate of 78 MHz, and a pump wavelength of 750 nm with a spot size of 1μm. The sample is placed in an optical cryostat, the light is collected via the same objective as the pump (100x magnification, numerical aperture 0.6) and detected by an InGaAs detector. In the following we shall compare the performance of purely photonic devices with metal-clad ones. In practice, as we will be looking at cavities which are comparable to or smaller than the spot size, this Gaussian distribution means that smaller cavities receive a proportionately larger fraction of the optical power, this is considered in simulations and the subsequent analysis.

Micro - photoluminescence (μ-PL) spectroscopy is performed at room temperature and at 100 K. Representative spectra of lasing devices are shown in Fig. 4(a) to 4(c), where we examined 500 nm wide square cavities without and with Au cladding at room temperature. Fig. 4(a) depicts the normalized PL spectra at the lasing onset and Fig. 4(b) and 4(c) show the power series for the purely photonic and the Au-clad case respectively. The peak at 900 nm in the PL spectra of the Au-clad device in Fig. 4(a) corresponds to the maximum PL emission of our InP material and does not correspond to a resonant mode emission. With increasing pump power, the emission peak of the cavity without Au first increases, then the device eventually degrades and is not lasing anymore, while for the Au-clad one the emission peak increases and saturates for higher pump power. For both devices, the emission peak shifts to shorter wavelengths for increasing pump powers. We attribute this characteristic blue shift to changes in the refractive index induced by free carrier absorption (plasma dispersion) which is known to be significant for InP [21, 41]. From the power series measurements, light-in-light-out (LL) curves as depicted in Fig 4(d) are constructed by extracting the maximum PL intensity for the resonant emission peaks. We expect an artificially broadened resonant peak due to the mentioned blueshift, that results in an increase of the full width at half maximum (FWHM) at high pump powers. This can be seen in Fig 4(e) where the FWHM of the resonant peak is shown for increasing pump powers. At first a linewidth narrowing is visible, but for higher excitation powers the FWHM increases again.

For metal-clad devices photonic modes can exist alongside plasmonic ones. To assess the plasmonic nature of a mode, one can consider the fraction of energy of a mode stored as kinetic energy of the free electrons [42], or the polarization of the mode. However, a clear distinction between a plasmonic and a photonic mode is difficult, unless - for example - certain geometrical effects [34] are employed to determine the polarization from the far field emission pattern. For the photonic modes, the metal shield introduces Joule losses but since the metal acts as a mirror, the confinement within the gain medium may be enhanced. As a consequence, the mode spreads less into free space in the metal-semiconductor-metal plane. [26] Extracted experimental $Q$ factors lie in the range of 100 - 300 for both type of devices. We expect to have scattering losses, for example due to the rough side walls of the devices and optical losses introduced by the metal. The PL emission at different excitation powers are captured with a standard camera. Images in Fig. 4(f) show emission below threshold, at the threshold and above threshold (from left to right) for a device without and with Au. Radiation patterns due to first order interference are distinctive for lasers and can be seen in both devices.

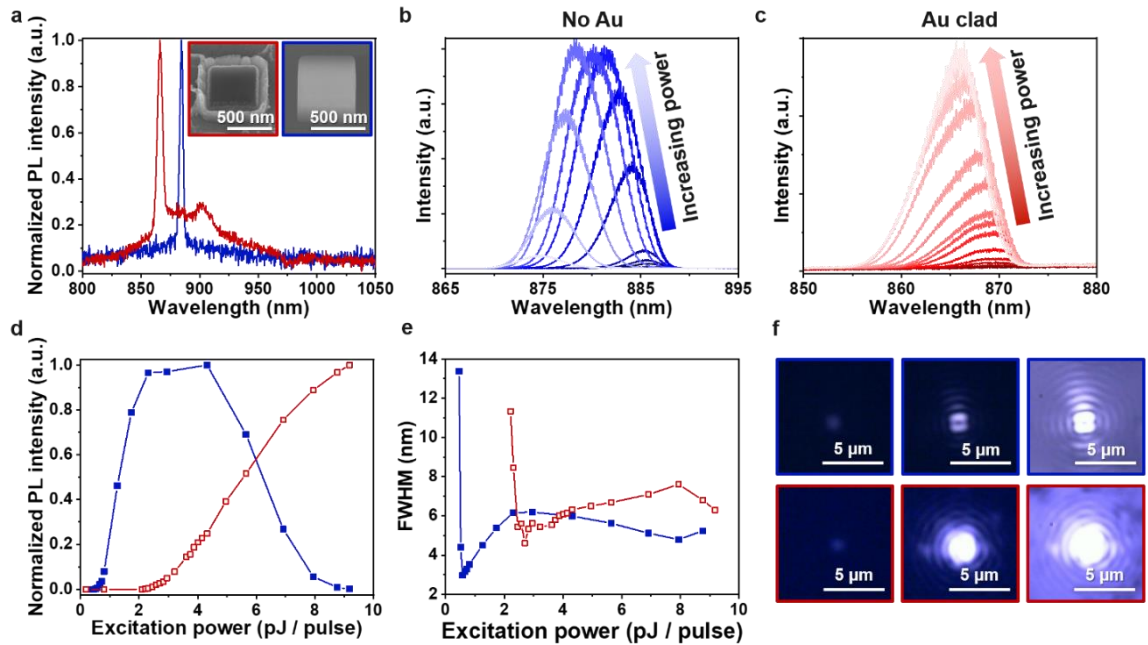

*Figure 4. (a) Spectra at lasing onset for 500 nm wide InP square devices without (blue) and with Au (red). Power series for resonant emission peak for the purely photonic (b) and the Au-clad (c) cavity. (d) Light-in light out curve for the devices and the corresponding FWHM extracted at different excitation powers (e). (f) Far field radiation pattern captured with a standard camera for photonic (top row, blue) and Au-clad (bottom row, red) cavities below threshold, at threshold and above threshold with the white background light on (from left to right). The measurements were done at room temperature. The lines in (d) and (e) are guides for the eyes.*

**Effect of cavity scaling**

Next, we explore the effect of scaling of the cavity dimension. It is important to remember that the Gaussian distribution of the pump source means that smaller cavities receive a proportionately larger fraction of the optical power. Also, in the following we will restrict the observations and discussions to the resonant emission with lowest threshold for the cases where a device exhibited multiple peaks. Fig. 5 shows the logarithmic (Fig. 5(a) and 5(b)) and linear (Fig. 5(c) and 5(d)) LL curves for different sized hexagonal cavities for the case without (Fig. 5(a) and 5(c)) and with (Fig. 5(b) and 5(d)) metal cladding. In case of multimode emission, we show the LL curves and the thresholds for the strongest emission peak only. One can observe three main differences between metal-cladded vs non- cladded structures:

First, the threshold is lower for the purely photonic devices. This is expected since the metal introduces optical loss, which must first be compensated before lasing sets in.

Second, the devices with a cross-section of less than 500 nm only lase in the metal-clad embodiment, but in this case the threshold is further increased compared to the larger metal-clad devices. For the larger (> 600 nm) devices the onset curve is steeper for the InP-only devices, indicating a lower threshold, and the saturation and degradation set in earlier than for the Au-clad devices.

Third, the purely photonic device saturates for lower pump powers and experiences an earlier and more dramatic drop-off in output power. Once a device starts to lase, it is generally expected that the increase in pump power generates a corresponding increase in laser output power (linear regime) and that the non-radiative decay which translates to heat generation remains more or less constant, until we reach saturation where the temperature should again start to increase leading to the reduction of output power. Based on the thermal simulations shown in Fig. 3, it is clear that the temperature in the cavity at threshold is substantially higher for the purely photonic devices. Hence, once the device enters saturation any additional temperature increase is expected to lead to a more abrupt reduction in lasing output power.

For larger devices the disadvantages of heating in a purely photonic cavity can be outbalanced by the reduced loss. However, when decreasing the diameter, radiative losses increase and combined with the large heating of these devices this means that lasing is no longer possible. The Au-clad devices show greater loss at lower excitation powers and larger gain cross section width because of the metal. When the devices are scaled, however, the metal provides increased confinement, independent of whether we are dealing with a plasmonic mode or just from improved reflection. At the same time the efficient heatsinking provided by the Au layer means that the cavity temperature is much less (about 40 - 50%) for the metal-clad device, so that these show lasing down to diameters of 300 nm in cross-section at 300 K.

Fig. 6 gives an overview of all the devices measured in terms of lasing threshold and emission versus diameter. In the case of Au-clad hexagons and squares we observe a larger variation of lasing threshold for devices with the same geometry by design. We believe that this might stem from slight variations in the cavity resulting from the Au milling process or the roughness of the Au grains. Consequently, several devices were measured to establish an average value for this comparison. The error bars indicate the standard deviation of the threshold. Furthermore, the different devices geometries will naturally result in slight deviations in active volume and in the number of modes which may exist in a device. However, as it may be observed in Fig. 6, the geometrical shape itself has little impact on the scaling potential, and we observe similar threshold trends regarding the device performance for all three geometries.

Similar to Fig. 5 we can observe an increased threshold for the metal-clad cavities, especially as we increase the dimensions, which would lead to an increased overlap of the optical mode with the metal. Nevertheless, it is interesting that despite the higher threshold we can scale the metal-clad cavities further down in diameter and maintain lasing. An interesting observation can be made when we observe the threshold dependence for 100 K (dotted lines in Fig. 6). The cavities without and with Au show similar thresholds, which are somewhat lower than the room temperature values for the purely photonic devices, but substantially reduced for the metal-clad devices. While the photonic cavities show resonant emission down to 500 nm the smallest metal-clad lasers now are only 150 nm in diameter. We expect that the cryogenic operation will mitigate the temperature increase caused by the pump laser, reduce the losses in the metal, decrease electron-phonon scattering at the metal-gain interface [43], and increase the achievable gain.

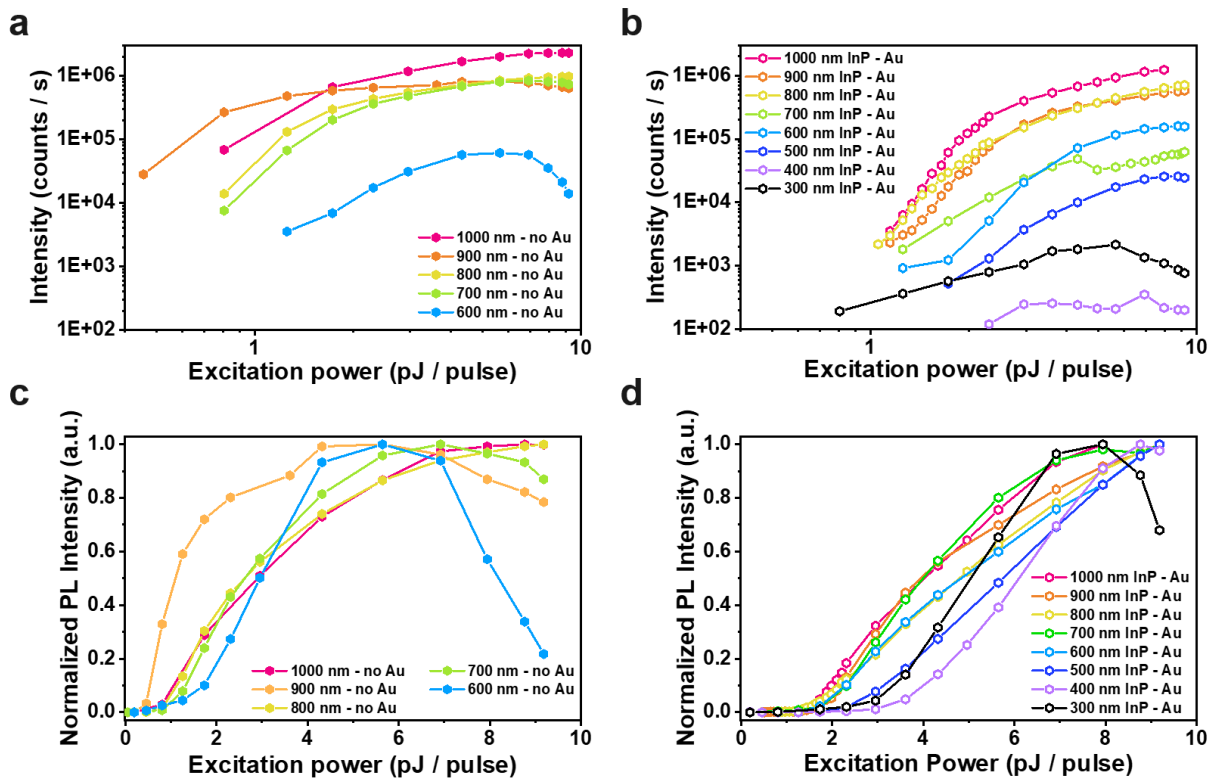

*Figure 5.* *Logarithmic LL curves for hexagonal cavities of varying diameter without (a) and with (b) Au cladding. Linear LL curves for hexagonal cavities (c) without and (d) with Au cladding used for determining the lasing threshold. The lines are guides for the eyes.*

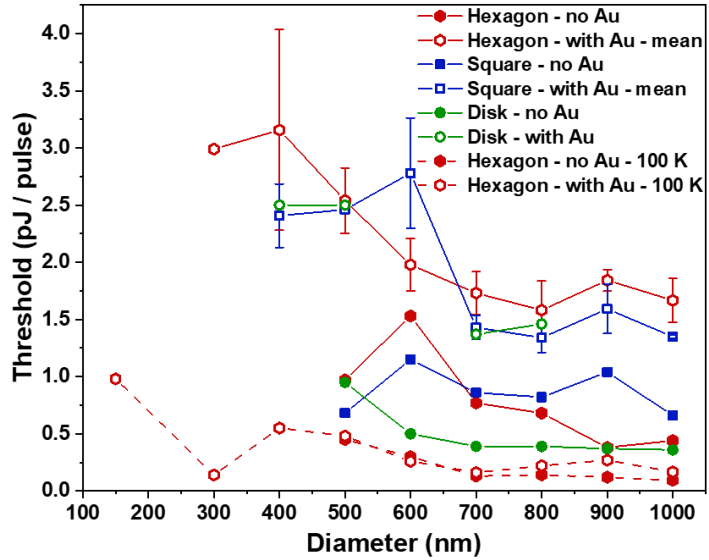

*Figure 6. Threshold versus diameter for devices with varying shape (hexagon (red), square (blue), disk (green)) and configurations without (filled symbols) and with (open symbols) Au cladding and the corresponding emission wavelength of the modes with the lowest threshold (b) at room temperature. The threshold values connected with the dashed line in (a) correspond to measurements of hexagonal cavities at 100 K. The lines are guides for the eyes.*

**Lifetime measurements**

Carrier lifetimes are determined by time-correlated-single-photon-counting (TCSPC). The optical response is detected by a Si single photon detector. For excitation, the same ps-pulsed laser is used as for the micro-PL spectroscopy and the lifetime is measured over the entire PL emission spectrum of InP. The resolution limit for the lifetime is around 40 ps, corresponding to the temporal pulse shape of the pumping laser. Lifetime measurements are performed at 100 K since different loss mechanisms are less significant at lower temperatures and hence, lead to overall longer lifetimes. This ensures a carrier lifetime longer than the resolution limit of the optical setup.

In Fig. 7(a) TCSPC measurements at 100 K are depicted for 600 nm wide hexagon cavities with Au cladding. The photon count rapidly decreases above threshold, indicating lasing. Fig. 7(b) shows normalized linear LL curves and lifetimes of 600 nm wide hexagons without and with Au cladding at different excitation powers. In both cases, the lifetime strongly decreases at the threshold and drops below our setup's resolution limit for higher pumping powers. In this case the lifetimes are similar with or without a metal cavity, but they drop off faster with increasing power for the metal-clad devices.

In Fig. 7(c) power dependent lifetimes are plotted for 300 nm, 600 nm, and 1000 nm wide hexagons with and without Au in addition to reference values measured outside the cavities. The cavities show a strong dependence of the lifetime on the excitation power and the smaller cavities exhibit a faster lifetime. This trend is stronger for the metal-clad devices, so that for the 1000 nm wide device the metal-clad cavity shows the longest lifetime at low pumping powers, whereas for the 300 nm wide cavity, the metal-clad device has a lifetime which is slightly less than the purely photonic devices. We interpret this to signify that for larger diameters where the modes in both the cavities with and without metal are expected to be photonic modes, the impact of the metal cladding is less significant in terms of increased loss as for the heat-sinking it provides, so that the increased temperature in the purely dielectric cavity results in a shorter lifetime.

For smaller cavities, we believe that the interaction with the metal becomes more significant as do radiative losses in the purely dielectric case. Hence, this results in overall shorter lifetimes, and the impact of increased temperature is less significant in comparison

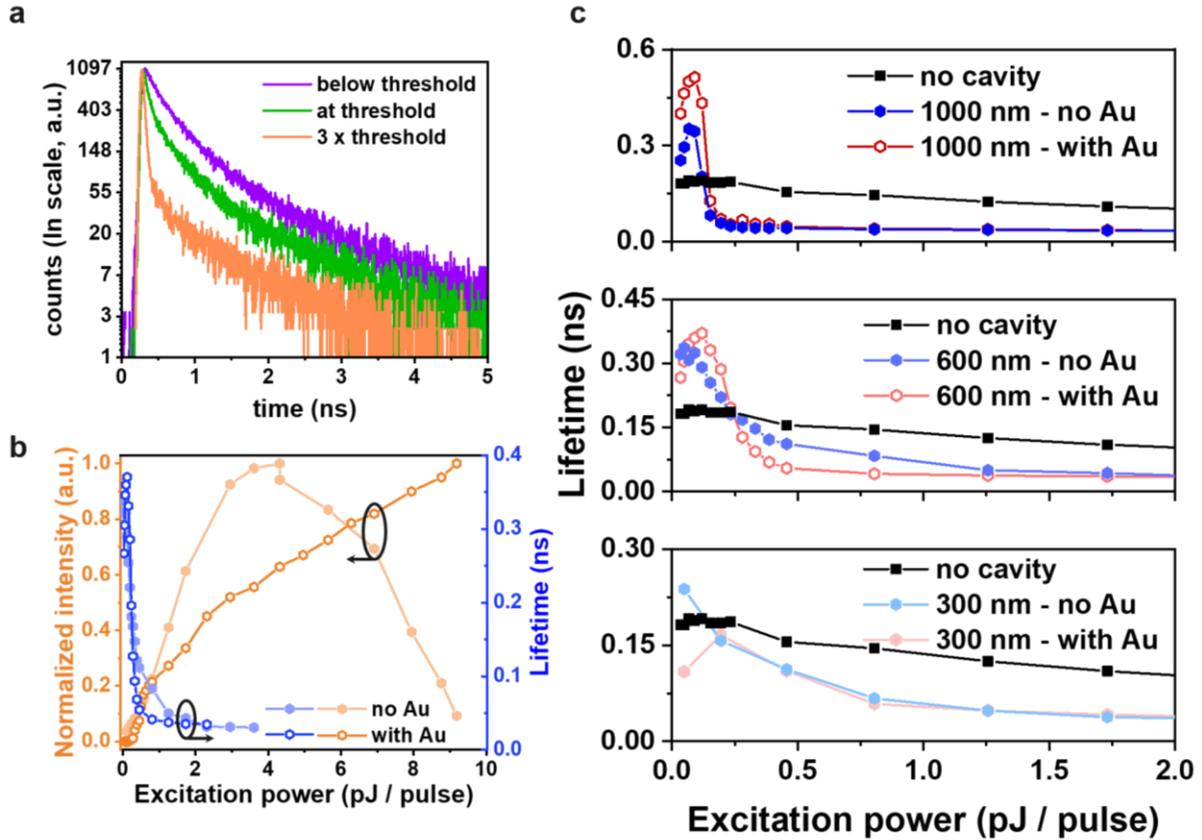

*Figure 7. (a) TCSPC measurements of a 600 nm wide Au-clad hexagon cavity at 100 K, below, at and above threshold. (b) Normalized linear LL curve for a 600 nm wide hexagon without (light orange, filled) and with (dark orange, empty) Au and corresponding lifetimes (light and dark blue) measured at 100 K. The lifetime sharply drops at the threshold and is longer for the device without gold. (c) Carrier lifetime at different excitation powers for 1000 nm, 600 nm and 300 nm wide Hexagon cavities without (blue) and with (red) Au cladding alongside reference values for bonded InP without a resonant mode (dashed line) at 100 K. In contrast to the reference values, there is a strong dependence of the lifetime on the pumping power.*

**Conclusion and Outlook**

Table 1 below summarizes our results of both experimental measurements and thermal simulations. In the smallest cavities which are 150 nm in width and possibly the 300 nm wide ones we most likely have hybrid photonic-plasmonic modes simply because these cavities are so small that a purely photonic mode would not be maintained based on consideration of the diffraction limit. For the intermediate sizes 400 – 600 nm, it is difficult to establish based on experimental evidence whether a mode is purely photonic or a hybrid mode, because the symmetric nature of our structures does not allow for a distinction based on the polarization of the light. However, independent of the nature of the mode it is possible to establish some general guidelines as to the impact of metal cavities based on our results: The thermal simulations show a dramatic increase in temperature within the cavity for the purely dielectric cavities, simply because the thermal transport through the thick dielectric capping it below and above is very inefficient. The temperature increase of up to 500 K (absolute temperature of about 800 K) simulated here are considered as a maximum as explained in the corresponding section, however it is impressive when you compare with the few tens of K increase observed in the metal-clad cavities. Hence, the excellent heat-sinking capability of the Au-cladding to a large extent mitigates the losses introduced by the metal. The threshold of purely dielectric cavities is still comparatively smaller than that of the metal-clad counterparts, on the other hand these show a more abrupt increase in the LL curve and a faster saturation and eventual degradation – most likely as a result of excessive heating of the nanolasers. When operating the lasers in a cryogenic environment (100 K) the overall losses will decrease both in the metal and the active materials and the gain increase. Hence, the difference between devices with and without a metal cladding is reduced. Although we still expect a similar temperature increase from the pumping, the impact will be smaller as absolute values of temperature are less. Stability and room temperature operation is one of the challenges of nanolasers, which are particularly susceptible to heating because of the large pump power or current densities needed in the small cavities, compared to conventional larger structures. The use of metal-clad cavities is therefore of interest as a design optimization, independent of whether the resulting mode is plasmonic or not in nature. The process to clad the devices with standard metal deposition techniques can be extended to monolithically integrated cavities as well, where one could benefit from atomically smooth sidewalls compared to the bonded devices [37] and it could be developed further for laterally contacted WGM lasers [44]. In this work we have carried out a thorough analysis of the operation of such devices using well-established processing methods which are not optimized for ultra-smooth metals. We therefore believe that this work will be of value in the design of novel nanolaser architectures.

| | Au | Device cross section diameter | | | | | | | | |
|---|---|---|---|---|---|---|---|---|---|---|
| | | 150 | 300 | 400 | 500 | 600 | 700 | 800 | 900 | 1000 |
| RT Threshold [pJ / pulse] (Square) | No | | | | 0.7 | 1.2 | 0.9 | 0.8 | 1 | 0.7 |
| | Yes | | 3.0* | 2.4 | 2.6 | 2.8 | 1.4 | 1.3 | 1.6 | 1.4 |
| 100 K Threshold [pJ / pulse] (Hexagon) | No | | | | 0.45 | 0.3 | 0.13 | 0.14 | 0.12 | 0.09 |
| | Yes | 0.98 | 0.14 | 0.55 | 0.48 | 0.26 | 0.16 | 0.22 | 0.27 | 0.17 |
| Maximum temperature [K] (Square) | No | | 587 | 667 | 730 | 775 | 796 | 805 | 805 | 798 |
| | Yes | | 314 | 315 | 317 | 318 | 320 | 322 | 324 | 325 |

*Table 1. Comparison of various metrics as a function of device cross-section diameter: Room-temperature (RT) threshold and maximum simulated temperature for square cavities with and without an Au-cladding, and lasing threshold for hexagon cavities with and without Au-cladding at 100 K. * Threshold for 300 nm wide device corresponds to hexagonal cavity.*

**Acknowledgment**

The authors gratefully acknowledge Markus Scherrer, Bernd Gotsmann, Heinz Schmid, and Andreas Schenk for fruitful technical discussions, as well as Yannick Baumgartner and the BRNC staff for technical support. The work presented here has received funding from the European Union H2020 ERC Starting Grant project PLASMIC (Grant Agreement No. 678567), H2020 MSCA IF project DATENE (Grant Agreement No. 844541), and the SNF Spark project SPILA (Contract No. 019_178).

**Author Contributions**

P.T., S.M., N.V.T, and K.M., created the device concept. P.W. performed Ansys simulations. P.T. performed Lumerical FDTD simulations, P.T. fabricated the sample with support of D.C. for the wire bonding. P.T performed the optical characterization on the devices with support from S.M. for the lifetime measurements. P.T., and N.V.T, analyzed the data. M.S. performed the FIB lamella and STEM and EDS characterization. K.M. lead and managed the project. All authors discussed the results. The manuscript was written by P.T, P.W. and K.M., with contributions of all authors, and all authors have given approval to the final version of the manuscript.